\documentclass{article}
\usepackage{graphicx} % Required for inserting images
\usepackage[utf8]{inputenc}
\usepackage[english]{babel}
\usepackage{csquotes}
\usepackage{amsmath}

\usepackage[
backend=biber,
style=alphabetic,
sorting=ynt
]{biblatex}

\title{Bias and Error Mitigation in Software-Generated Data: An Advanced Search and Optimization Framework Leveraging Generative Code Models}
\author{Ernesto Giralt Hernández }
\date{Barcelona, October 2023}

\begin{document}

\maketitle

\begin{abstract}
Data generation and analysis is a fundamental aspect of many industries and disciplines, from strategic decision making in business to research in the physical and social sciences. However, data generated using software and algorithms can be subject to biases and errors. These can be due to problems with the original software, default settings that do not align with the specific needs of the situation, or even deeper problems with the underlying theories and models. 
This paper proposes an advanced search and optimization framework aimed at generating and choosing optimal source code capable of correcting errors and biases from previous versions to address typical problems in software systems specializing in data analysis and generation, especially those in the corporate and data science world. Applying this framework multiple times on the same software system would incrementally improve the quality of the output results. It uses Solomonoff Induction as a sound theoretical basis, extending it with Kolmogorov Conditional Complexity, a novel adaptation, to evaluate a set of candidate programs. We propose the use of generative models for the creation of this set of programs, with special emphasis on the capabilities of Large Language Models (LLMs) to generate high quality code. 
\end{abstract}


\newpage
\tableofcontents
\newpage

\section{Introduction}
The production and scrutiny of data underpins a wide range of sectors and academic fields, from high-stakes decision making in business to multifaceted research in the natural and social sciences. However, data from software or algorithmic processes are not immune to biases and errors. These inaccuracies can come from a variety of sources, such as problems in the initial software, misaligned default settings, or even fundamental flaws in the theories and models on which the software is built.

In corporate environments, where data accuracy and reliability are non-negotiable, the presence of biases and errors in software-generated data can have far-reaching and sometimes serious consequences. Traditional debugging and error correction techniques can fall short in identifying and rectifying these problems, especially when dealing with complex algorithms and large data sets.

This paper presents a novel theoretical framework tailored to advanced optimization and search for software-generated data. Unlike existing models to date, which may be computationally cumbersome or inapplicable to practical real-world datasets\cite{hutter2007universal}, the framework we propose is computationally feasible and practically applicable.

The unique strength of this approach lies not just in the consideration of a finite set of program alternatives, encapsulated in the set \( \mathcal{P} \), but also in the explicit definition of finite and well-structured input \( \Theta \) and output datasets \( D \) and \( D^* \) \cite{paulmb2020vi}. Our model goes beyond merely applying Bayesian inference and Solomonoff Induction by incorporating additional components such as a specific likelihood function and a redefined utility function. These elements enable our framework to be highly effective, computationally efficient, and directly applicable to the problem under consideration. 

This comprehensive methodology allows us to assess how each program variant in \( \mathcal{P} \) fits the observed data \( D \), its inherent complexity, and the extent of bias and error it may introduce into the generated data. This yields a harmonious trade-off between simplicity and accuracy, empowering the model to select the program variant that most effectively minimizes biases and errors while maintaining computational efficiency.

By focusing on domains defined by datasets generated through known programming languages such as Python, we restrict our scope to render the model practically implementable without sacrificing the core principles that inform this advanced optimization and search framework. 

The introduction of this innovative framework opens up promising avenues for significantly enhancing the reliability and robustness of data generated by software systems, with potential ramifications across diverse sectors from financial systems to scientific research.

\section{Mathematical Symbols}
In this article, we utilize various mathematical symbols to represent different concepts and calculations. Here is an explanation of each symbol used in the equations:

\begin{itemize}
    \item \(D\): Represents the observed (output) data set.
    \item \(\Theta\): Denotes the input parameter space.
    \item \(\mathcal{P}\): Refers to the set of programs \(p\) or variants under consideration.
    \item \(p\): Represents an individual program or variant within the set \(\mathcal{P}\).
    \item \(p^*\): Refers to the objective optimal program that maximizes the utility function.
    \item \(MSE(p, D)\): Represents the Mean Squared Error between the generated data set by program \(p\) and the observed data set \(D\).
    \item \(MLE\) : Maximum Likelihood Estimation.
    \item \(L(p|D)\): Denotes the likelihood function, which measures the probability of program \(p\) given the observed data set \(D\).
    \item \(U(p)\): Represents the utility function, which evaluates the overall value of a program \(p\) based on fit to the data and other criteria.
    \item \(P(p|\Theta, D)\): Represents the posterior probability of a program \(p\) with a \(\Theta\) input dataset, given the observed data set \(D\).
    \item     \( R(p) \): Penalty function acting on each program in set \( P = {p_1, p_2,..., p_n} \), assigning a penalty to these programs.
    \item   \( \lambda \): A weighting factor affecting the balance between the posterior probability and penalty function within the utility function for program \( p \).
\end{itemize}

\newpage
\section{Domain Specification for the Adapted Model: Finiteness, Decidability, and Computability}

In this work, we examine a model that operates within a highly specific and constrained domain, characterized by its finiteness, decidability, and computability. We introduce a set \( D \) of observations generated by a collection of programs \( \mathcal{P} = \{p_1, p_2, \ldots, p_n\} \). A different set \( \Theta \) serves as the input parameter space for these programs. Among these, an optimal program \( p^* \) is identified, capable of generating an optimal dataset \( D^* \).

\textbf{Observation Set \( D \) and Optimal Observation \( D^* \)}: 
Both \( D \) and \( D^* \) are finite sets due to their bounded number of columns (variables) and rows (cases). The practical limitations of data collection, storage, and processing ensure this finitude.

\[
D = \{d_1, d_2, \ldots, d_m\}
\]
\[
D^* = \{d_1^*, d_2^*, \ldots, d_m^*\}
\]

\( D \) and \( D^* \), being finite sets, are computable. They can be explicitly enumerated and manipulated. \(D^*\) comes from an optimal prediction of the set \( D \) from the result of a previous evaluation of \( D \) that has determined biases or errors. This previous evaluation could be, for example, an anomaly analysis that has determined problems in \( D \). The operation and definition of this evaluation is beyond the scope of this framework, but it is the indirect origin of \(D^*\).

Due to their finite and well-defined nature, any property or computational task involving \( D \) and \( D^* \) can be algorithmically determined.

\textbf{Program Space \( \mathcal{P} \)}:
The collection \( \mathcal{P} \) could be generated by a model \( G(p|\Theta, D) \)\cite{brown2020language}, initiating from a base program \( p \) and creating a finite set of programmatic variations. Given that \( G \) operates within set computational and memory constraints, it ensures that \( \mathcal{P} \) remains finite.

\[
\mathcal{P} = \{p_1, p_2, \ldots, p_n\}
\]
\[
p^* = \arg\max_{p \in \mathcal{P}} U(p)
\]

The space \( \mathcal{P} \) is both finite and decidable, allowing for efficient identification of \( p^* \) via traversal of \( \mathcal{P} \). The finitude of \( \mathcal{P} \) allows for algorithmic determination of \( p^* \), the optimal program for generating \( D^* \), given a specific utility function \( U(p) \).
\[
D^* = p^*(\Theta)
\]

\section{Generative Approaches of Program Variants in Finite Set \( \mathcal{P} \)}
The generation of program variants \( p \) capable of producing an optimal dataset \( D^* \) can be effectively achieved through Large Language Models (LLMs) such as GPT-3\cite{openai2020gpt3}, GPT-4\cite{openai2023gpt4}, LlaMa-2\cite{meta2023llama2}, among others. These models have demonstrated the ability to produce high-quality, syntactically correct, and semantically coherent code. Alternative methods for generating program variants are also available but may lack the level of automation and quality that LLMs offer.

The effectiveness of code generation can be further enhanced by utilizing fine-tuned versions of these models. Commercial offerings like Github Co-Pilot\cite{github2022copilot} serve as examples. Fine-tuned models, when pre-trained on the original source code, can produce variants with higher degrees of completeness and accuracy.

\subsection{Parameterization with Optimal Dataset \( D^* \)}
The presence of an optimal dataset \( D^* \) serves as a guiding objective for the generative model. This dataset can be used as a pre-condition or parameter during the generation phase. Advanced parameterization techniques may also incorporate elements from the penalty function \( R(p) \) and the weighting factor \( \lambda \), thereby allowing for a dynamically configurable generative model.

On the other hand, the utility function \( U(p) \) plays a pivotal role in this framework by ensuring the selection of the optimal program variant from \( \mathcal{P} \). This function is especially crucial when multiple generative methods are used, offering a standardized metric for comparison and selection.

\subsection{Considerations on Variant Quantity}
The number of program variants generated by the model can be a tunable parameter in the framework. Several approaches can be considered:
\begin{enumerate}
    \item A fixed number of variants could be generated per iteration, allowing for systematic evaluation.
    \item The number of variants could be dynamically determined based on the performance of previously generated programs, as quantified by \( U(p) \).
    \item A hybrid approach could initially generate a large pool of variants, followed by more targeted generation based on utility function feedback.
\end{enumerate}

\subsection{Future Enhancements}
As the framework matures, one could consider implementing a localized learning model that trains on configurations of past successful programs. This would enable the generative model to progressively improve the quality of the generated program variants.

\section{Incorporating Conditional Kolmogorov Complexity into Solomonoff Induction for Program Evaluation}

\subsection{Conditional Kolmogorov Complexity}
Kolmogorov complexity, in its traditional form, quantifies the amount of computational resources needed to specify a string \( x \). For a given string \( x \), its Kolmogorov complexity \( K(x) \) is defined as the length of the shortest Turing machine program \( p \) that produces \( x \) when run. Essentially, \( K(x) \) measures the inherent complexity or randomness contained within \( x \). \cite{li2008kolmogorov}.

In formal terms, the Kolmogorov complexity \( K(x) \) of a string \( x \) is defined as:
\[
K(x) = \min_{p:U(p)=x} |p|
\]
where \( U \) represents a universal Turing machine, \( p \) is a program that, when run on \( U \), generates \( x \), and \( |p| \) is the length of \( p \).

In certain applications, it may be beneficial to evaluate a program \( p \) in the context of another program \( p' \), particularly when \( p \) is a variant or modification of \( p' \).  In the context of source code comparison, for example, we propose an adaptation of this complexity measure that aligns well with the practical requirements of software development. Source code, though semantically rich, can be converted into a binary string devoid of semantic meaning (Shannon wrote “ ... semantic aspects of communication are irrelevant to the engineering problem ..”)\cite{shannon1948mathematical}  thereby becoming a suitable subject for complexity analysis.

As an example, Git, a version control system, could be used to compare two versions of the same source code. Git provides a 'diff', essentially a list of changes between two versions. This diff captures the textual differences at the granularity of lines or even individual characters and can be treated as a binary string.\cite{chacon2014pro}

To approximate the Kolmogorov complexity \( K(p) \) of a given source code \( p \), we propose measuring the length (in bytes) of the diff generated when comparing \( p \) with a reference version \( p' \). This length serves as an estimation of \( K(p|p') \), the conditional Kolmogorov complexity of \( p \) given \( p' \).\cite{fouad2017} 

In this way, we can state that one version of the source code is "shorter" than another with respect to the original program by considering the byte length of their respective diffs. This adapted measure offers a practical yet theoretically grounded method for assessing the complexity of source code versions relative to each other.

Considering all this, we introduce the concept of \textbf{conditional Kolmogorov complexity}, denoted as \( K(p|p') \), - replacing \(p\)  as the string \( x \) -  which measures the complexity of \( p \) given \( p' \). Formally, it is defined as:

\[
K(p|p') = \min_{U(q, p') = p} |q|
\]
This measure captures the "cost" in terms of complexity to transform \( p' \) into \( p \).

\subsection{Adapted Solomonoff Induction}
Solomonoff's theory of inductive inference provides a formal foundation for predicting future data based on observed data\cite{rathmanner2011philosophical}. It does so by using the concept of Kolmogorov complexity, which quantifies the amount of computational resources needed to specify an object, blending algorithmic information theory with Bayesian probability. The traditional formulation for the probability \( P(x) \) of observing a string \( x \) is expressed as:
\[ P(x) = \sum_{p:U(p)=x} 2^{-|p|} \]
where \( U \) is a universal Turing machine, \( p \) is a program that produces \( x \) when run on \( U \), and \( |p| \) is the length of \( p \) (Kolmogorov complexity of \(p\)). 

Applying our specific domain  formulation to calculate the posterior probability \( P(p|\Theta,D) \) of a program \( p \) given an input \( \Theta \) and data \( D \), the new formulation is as follows:

\[
P(p|\Theta,D) = \frac{2^{-K(p)} \times L(p|D)}{\sum_{p' \in \mathcal{P}} 2^{-K(p')} \times L(p'|D)}
\]

Here, \( K(p) \) is the Kolmogorov complexity of the program \( p \), and \( L(p|D) \) is the likelihood function evaluating how well the program fits the data \( D \).

The inclusion of conditional Kolmogorov complexity leads us to a modified version of the posterior probability in Solomonoff Induction:
\[
P(p|\Theta,D) = \frac{2^{-K(p|p')} \times L(p|D)}{\sum_{p' \in \mathcal{P}} 2^{-K(p'|p)} \times L(p'|D)}
\]

\section{Advanced Utility Function}
The traditional formulation of Solomonoff Induction is based solely on algorithmic probability\cite{solomonoff1964formal}. That is, the probability of an event (or hypothesis) is inversely proportional to the length of the shortest program that generates it. In a universal sense, it can be used to make predictions about future data points by summing over all possible programs weighted by their complexity.

For our approach the utility function \( U(p) \) has been re-defined as:
\[ U(p) = \lambda P(p|\Theta,D) + (1-\lambda)R(p) \]
Where \(R(p)\) is a penalty function and \(\lambda\) is a weighting factor.

It introduces a nuanced approach to decision-making in the context of Solomonoff Induction. By incorporating this utility function adapted with this new components, the modified Solomonoff Induction model can be more applicable in practical scenarios, especially when dealing with a computable and finite set of Turing machines, as in the case of model-driven or model-generated software algorithms. The dual consideration of algorithmic probability and a penalty function allows for a more flexible and nuanced approach to inductive reasoning and decision-making\cite{hutter2013algorithmic}. This enhanced framework has the potential to be valuable in fields where traditional Solomonoff Induction might be too idealistic or computationally infeasible.

The utility function \( U(p) \) is designed to evaluate the aptitude of each program \( p \) in \( \mathcal{P} \) for generating an optimal set \( D* \)  that maximizes this utility. It has been defined is a weighted combination of two terms:
\begin{enumerate}
    \item \( P(p|\Theta,D) \): This is the posterior probability of a program \( p \) given an input \( \Theta \) and data \( D \). This can be interpreted as a measure of how likely a given program \( p \) is, given the available evidence. The higher this probability, the more "reliable" or "believable" the program is considered to be.
    \item \( R(p) \): This is a penalty function that assigns a penalty to different programs. In the context of a decision-making agent, this could be interpreted as a cost function. The lower the value of \( R(p) \) for a given program \( p \), the higher the penalty associated with that program.
\end{enumerate}

\subsection{Penalty function}
 A crucial modification was the introduction of a penalty term \(R(p)\) in the utility function. This term can be engineered to capture various facets of the program, such as computational complexity, robustness, or even ethical considerations. The penalty term makes the utility function differentiable, allowing for the use of advanced optimization techniques like gradient-based methods or even quantum optimization in future implementations. 
 
On the other hand, the factor \( \lambda \) is a weight that allows for balancing the relative importance between the posterior probability and the penalty function. When \( \lambda = 1 \), the decision is solely based on the posterior probability \( P(p|\Theta,D) \) and completely disregards the penalty function. Conversely, when \( \lambda = 0 \), only the penalty function \( R(p) \) is considered and the posterior probability is ignored.

To determine an optimal value for \( \lambda \) in this context:
\begin{enumerate}
    \item \textbf{Consider the confidence in the data}: If you have high confidence in the accuracy and completeness of \( \Theta \) and \( D \), you might want to give more weight to \( P(p|\Theta,D) \) by increasing the value of \( \lambda \). In contrast, if the data is uncertain or noisy, you might want to give more weight to the penalty function.
    \item \textbf{Consider the severity of the penalties}: If the penalties in \( R(p) \) are particularly severe or critical to the problem (e.g., if some programs in \( P \) are absolutely unacceptable for some reason), you might want to give more weight to \( R(p) \) by decreasing the value of \( \lambda \).
    \item \textbf{Empirical tests}: If possible, you might conduct a series of empirical tests with different values of \( \lambda \) to determine which produces results more aligned with the problem's expectations or objectives.
    \item \textbf{Domain knowledge}: If there is prior knowledge about the relationship between probability and penalties, this knowledge could be used to inform the choice of \( \lambda \).
\end{enumerate}

Finally, one strategy might be to start with \( \lambda = 0.5 \) to give equal weight to both components and adjust as necessary based on observed results and the above considerations. It's essential that any choice of \( \lambda \) be adequately justified based on the problem's characteristics and priorities.

\subsubsection{How to define the Penalty Function}
Given the nature of the problem —finding a program \(p\) capable of producing an optimal observation \(D^*\)— the penalty function \(R(p)\) should be designed to penalize deviations of the observation \(D\) with respect to \(D^*\). Considering this, there are several ways in which \(R(p)\) could be defined:

\begin{enumerate}
    \item \textbf{Quadratic Penalty}:
\[ R(p) = \alpha (D - D^*)^2 \]
Where \(\alpha\) is a scaling factor. This penalty more severely penalizes large deviations from \(D^*\) and is less harsh with small deviations.

\item \textbf{Absolute Penalty}:
\[ R(p) = \alpha |D - D^*| \]
Similar to the quadratic penalty, but it penalizes deviations linearly.

\item \textbf{Threshold-based Penalty}:
\[ 
R(p) = 
\begin{cases} 
0, & \text{if } |D - D^*| < \epsilon \\
\alpha, & \text{if } |D - D^*| \geq \epsilon 
\end{cases}
\]
This penalty does not penalize observations within a margin \(\epsilon\) of \(D^*\), but applies a penalty \(\alpha\) to observations that deviate beyond that margin.

\item \textbf{Complexity-based Penalty}: Observations could be penalized if they are generated from unnecessarily complex programs, favoring simpler and more elegant solutions. For example:
\[ R(p) = \alpha L(p) \]
Where \(L(p)\) is a measure of the length or complexity of \(p\).
\end{enumerate}

The optimal choice of the penalty function depends on specific goals and constraints:

\begin{itemize}
    \item If the aim is for observations to be as close as possible to \(D^*\), a quadratic penalty might be suitable.
    \item If small deviations from \(D*\) are tolerable, but large deviations are not, the threshold-based penalty would be ideal.
    \item If the goal is to find simple and elegant solutions, the complexity-based penalty might be most appropriate.
\end{itemize}

The key is to align the penalty function with specific goals for the problem. It might also be beneficial to experiment with various penalty functions and assess which one produces the best results for the specific use case.

\subsection{Likelihood Function Design and Bias Robustness}
The design of the likelihood function \(L(p|D)\) adopts a sigmoidal form to ensure several key properties. Firstly, it bounds the influence of extreme MSE values. This is especially critical when dealing with software-generated data that might be prone to outliers or biases. By constraining the impact of extreme MSE values, the function offers enhanced robustness.\cite{bishop1995neural}

Secondly, the sigmoidal function inherently acts as an activation function. This introduces a degree of non-linearity into the model. Such non-linearity enables the model to capture intricate relationships within the data that might be missed with a purely linear approach.

The probability of a program \(p\) producing a given observation  \(D\) based on input \(\Theta\) is defined by:
\[ P(p|\Theta,D) = \frac{\text{MLE}(p|D) \times L(p|D)} {\sum_{p' \in \mathcal{P}} L(p'|D)} \]
In this equation, \(\text{MLE}(p|D)\) represents the Maximum Likelihood Estimation for \(P(p|D)\).  It is a standard approach in statistics, aiding in identifying the parameter value that maximizes the likelihood function. \cite{hastie2009elements}.  

The likelihood function \(L(p|D)\) is described by:
\[ L(p|D) = \frac{1}{1 + e^{-MSE(p, D)}} \]

This function employs a sigmoid to transform the domain of \(MSE(p, D)\) from all real numbers to a range between 0 and 1. This transformation is critical in preventing extremely high or low \(MSE(p, D)\) values from exerting a disproportionate influence on the likelihood function.

For normalization, the function divides by \(\sum_{p' \in \mathcal{P}} L(p'|D)\). This ensures that the probabilities across all possible programs in set \( \mathcal{P} \) sum to 1, a necessary condition for producing valid probabilities.

In the context of evaluating programs \(p\) from the entire set \(P\) capable of generating an observation \(D\) , this approach offers a methodical way to determine how the given observation deviates from the optimal observation \(D^*\). 

\subsubsection{Maximum Likelihood Estimation (MLE)}
MLE is a method used for estimating the parameters \( \theta \) of a statistical model \cite{casella2002statistical}. Given a set of observations \( D \), MLE aims to find the parameter values that maximize the likelihood function \( L(\theta|D) \):
\[
\theta^* = \arg\max_{\theta} L(\theta|D) = \arg\max_{\theta} \prod_{i=1}^{N} P(x_i|\theta)
\]
Here, \( \theta^* \) is the parameter value that maximizes the likelihood of observing the given data \( D \) under the model parameterized by \( \theta \).

\subsection{Optimal Prediction and Optimal Decision in Solomonoff Context}
The optimal prediction for a new observation \(x\) is computed as a weighted average over all programs in \(\mathcal{P}\), using the posterior probabilities as weights \cite{bishop2006pattern}:
\begin{equation}
P(x|D) = \sum_{p \in \mathcal{P}} P(x|p, D) \times (P(p|\Theta, D))
\end{equation}
In this adapted model, the optimal decision \(p^*\) is taken as the program with the highest posterior probability:
\begin{equation}
p^* = \arg\max_{p \in \mathcal{P}} P(p|\Theta, D)
\end{equation}
The equation essentially encapsulates the primary objective of the model: to find the program that yields the highest utility.

\section{Integration of the Likelihood Function and Advanced Utility Function for Optimal Program Selection}
In the effort to evaluate and select the most suitable program from a set, two distinct but potentially complementary approaches have been proposed: the likelihood function \(L(p|D)\) and the fitted utility function \(U(p)\). Each of these functions serves different purposes, but when applied together, they provide a more holistic measure of a program's appropriateness. Below, we explain the role of both functions and justify their simultaneous application in our model evaluation framework.

\subsubsection{The Likelihood Function \(L(p|D)\)}
The likelihood function gauges how well a given program \(p\) fits the data \(D\), particularly in terms of its prediction error (MSE).  The function inherently provides a measure of the "fit quality" of the program to the dataset. It quantifies the closeness of the program's output to the observed data, emphasizing the accuracy of predictions.

\subsubsection{The Adjusted Utility Function \(U(p)\)}
The adjusted utility function measures the overall value of a program \(p\). It is not solely based on the fit to the data, but also integrates other valuable criteria like program simplicity, computational costs, or any other factor penalized by \(R(p)\). By combining the adjusted probability of the program with a penalty function, it facilitates a comprehensive valuation of the program, balancing fit and other user-defined criteria.

\subsection{Synergy Between Functions}
While the likelihood function emphasizes fitting accuracy, the utility function broadens the scope to consider a wider range of factors. Their combined application offers several advantages:
\begin{itemize}
    \item \textbf{Holistic Evaluation}: While the likelihood function narrows down programs based on fit, the utility function further refines this selection by evaluating broader criteria. This ensures that the chosen program is not only accurate but also optimal in other regards.
    \item \textbf{Robustness}: Using both functions successively could bolster the model against potential overfitting. While a program with a perfect likelihood might be too tailored to the training data, the utility function can penalize it if it's overly complex or not generalizable.
    \item \textbf{Flexibility}: Incorporating both functions allows for model adaptability. Depending on the application or domain, the importance of fit versus other factors can vary. Having both functions in the evaluation framework allows for such adaptability.
\end{itemize}

In summary, the dual-application of both the likelihood and utility functions provides a comprehensive, adaptable, and robust method for evaluating programs in set \(P\). By embracing this approach, we ensure that our model selection is not just data-driven but also aligned with broader objectives and constraints.

\section{Two-Phase Approach for Optimal Program Discovery}
The general algorithm for optimal program discovery can be broadly categorized into two major phases: the Preparation Phase and the Optimization Phase. Each phase has specific tasks that contribute to the overall objective of identifying an optimal program \( p^* \) capable of generating an optimal dataset \( D^* \).

\subsection{Preparation Phase}
\begin{enumerate}
\item \textbf{Generation of \( D \)}: The original program \( p \) is executed to produce a dataset \( D \), which becomes the initial point of reference for all subsequent optimization tasks.
\item \textbf{Evaluation of \( D \)}: Anomalies, biases, or other deficiencies in \( D \) are examined through various evaluation methods. The specifics of these methods can include statistical tests, expert reviews, or machine learning models trained to identify data anomalies. 
\item \textbf{Generation of \( D^* \)}: Based on the evaluation of \( D \), an optimal dataset \( D^* \) is created. This dataset acts as the `ground truth' and serves as the target output for all programs in the optimization phase. \( D^* \) is generated by applying corrections to the identified biases or errors in \( D \).
\item \textbf{Generation of \( \mathcal{P} \)}: A generative model, such as GPT-4, is employed to produce a finite set of programmatic variations \( \mathcal{P} \) based on \( p \). Each program in \( \mathcal{P} \) aims to generate a dataset that approximates \( D^* \) as closely as possible. 
\end{enumerate}

\subsection{Optimization Phase}
The Optimization Phase is where the actual search for the optimal program \( p^* \) occurs, guided by a utility function \( U(p) \) and a likelihood function \( L(p|D) \).

\begin{enumerate}
    \item \textbf{Initialization}:
    \begin{itemize}
        \item Choose an initial program \( p_{\text{init}} \) from \( \mathcal{P} \).
        \item Calculate \( L(p_{\text{init}}|D) \) and \( P(p_{\text{init}}|\Theta,D) \).
        \item Evaluate \( U(p_{\text{init}}) \) using \( P(p_{\text{init}}|\Theta,D) \), \( R(p) \), and \( \lambda \).
    \end{itemize}
    \item \textbf{Local Exploration}:
    \begin{itemize}
        \item For each program \( p_i \) in the neighborhood of \( p_{\text{init}} \):
        \begin{itemize}
            \item Calculate \( L(p_i|D) \) and \( P(p_i|\Theta,D) \).
            \item Evaluate \( U(p_i) \) using \( P(p_i|\Theta,D) \), \( R(p) \), and \( \lambda \).
        \end{itemize}
    \end{itemize}
    \item \textbf{Selection}:
    \begin{itemize}
        \item Choose \( p_{\text{new}} \) with the highest utility in the neighborhood.
    \end{itemize}
    \item \textbf{Update}:
    \begin{itemize}
        \item If \( U(p_{\text{new}}) > U(p_{\text{init}}) \), set \( p_{\text{new}} \) as \( p_{\text{init}} \) for the next iteration.
    \end{itemize}
    \item \textbf{Convergence}:
    \begin{itemize}
        \item Stop the algorithm if there is no improvement or if a maximum number of iterations is reached.
    \end{itemize}
    \item \textbf{Return}:
    \begin{itemize}
        \item \( p^* \) is the program with the highest utility found.
    \end{itemize}
\end{enumerate}

In each iteration, the utility calculation \( U(p) \) for a given program \( p \) involves both \( P(p|\Theta,D) \) and \( R(p) \). Furthermore, \( P(p|\Theta,D) \) itself incorporates \( L(p|D) \). In this way, both the likelihood function and the utility function are intrinsically involved in the evaluation of each program during the search process.

This two-phase approach provides a systematic and quantifiable method to improve upon an existing program \( p \), based on the optimization of a target dataset \( D^* \).

\section{Computational Complexity}
The computational complexity of our framework is an essential aspect that demands thorough evaluation. We consider the following components:

\begin{enumerate}
    \item \textbf{Traversal of Program Space}: The algorithm iteratively searches through the program space \( P \), which has cardinality \( |P| \). Given that the search algorithm is linear and local, this contributes \( O(|P|) \) to the overall complexity.

       \item \textbf{Likelihood Function}: For each program \( p \), we compute the likelihood \( L(p|D) \) as \( \frac{1}{1 + e^{-\text{MSE}(p, D)}} \). Since \( n \) is the size of the dataset \( D \), this operation is \( O(n) \).

    \item \textbf{Conditional Kolmogorov Complexity}: For each pair \( (p, p') \), we compute \( K(p|p') \) as \( \min_{U(q, p') = p} |q| \). Assuming that we can approximate this in constant time for each pair, it adds \( O(|P|) \) complexity.
\end{enumerate}

The overall complexity of the algorithm can thus be approximated as \( O(|P| \times n) \).

\newpage

\printbibliography[
heading=bibintoc,
title={Bibliography}
]

\end{document}